\documentclass[nobibnotes,nofootinbib,aps,pre,showpacs, preprint 
]{revtex4}
\usepackage{setspace}
\usepackage{graphics}
\usepackage{graphicx}
\usepackage{color}
\usepackage{amsmath}
\usepackage{amssymb}
\usepackage{bm}
\usepackage{float}
\usepackage{epsfig}
\usepackage{epstopdf}
\usepackage{color,graphics}
\usepackage[toc,page]{appendix}
\def\mod{\color{blue}}

\usepackage{array,multirow,makecell}
\usepackage{amsfonts}
\begin{document}
\title{Re\-la\-ti\-vis\-tic Ermakov-Milne-Pinney Systems and First Integrals} 

\author{Fernando Haas}
\affiliation{Physics Institute, Federal University of Rio Grande do Sul, Av. Bento Gon\c{c}alves 9500, 91501-970 Porto Alegre, RS, Brazil}

\begin{abstract}
The Ermakov-Milne-Pinney equation is ubiquitous in many areas of physics that have an explicit time-dependence,  including quantum systems with time dependent Hamiltonian, cosmology, time-dependent harmonic oscillators, accelerator dynamics, etc. The Eliezer and Gray physical interpretation of the Ermakov-Lewis invariant is applied as a guiding principle for the derivation of the special relativistic analog of the Ermakov-Milne-Pinney equation and associated first integral. The special relativistic extension of the Ray-Reid system and invariant is obtained. General properties of the relativistic Ermakov-Milne-Pinney are analyzed. The conservative case of the relativistic Ermakov-Milne-Pinney equation is described in terms of a pseudo-potential, reducing the problem to an effective Newtonian form. The non-relativistic limit is considered as well. A relativistic nonlinear superposition law for relativistic Ermakov systems is identified. The generalized 
Ermakov-Milne-Pinney equation has additional nonlinearities, due to the relativistic effects.
\end{abstract}

\maketitle

\section{Introduction}

The Ermakov-Milne-Pinney (EMP) equation \cite{Ermakov, Milne, Pinney}
\begin{equation}
\ddot{x}+\kappa^{2}(t)x = C/x^3 \,,
\end{equation}
where $C$ is a real constant usually taken as positive, is { an ubiquitous} nonlinear non-au\-to\-no\-mous ordinary differential equation with many applications, in particular in problems related to the time-dependent harmonic oscillator or in connection with exact solutions of the one-dimensional time-independent Schr\"odinger equation. In more generality, applications of the EMP equation appear in cosmological models 
\cite{Hawkins, Rosu, Riemann}, Bose-Einstein condensates \cite{Haas1, Herring}, photonic lattices \cite{Rod}, accelerator dynamics \cite{Courant, Qin}, gravitational wave propagation \cite{Ger}, higher order spin models \cite{Fring}, quantum plasmas \cite{EPL}, limit cycles \cite{Llibre}, dynamical symmetries \cite{Haas3}, magneto-gasdynamics \cite{Rogers}, { time-de\-pen\-dent non-Hermitian quantum system \cite{Dey, Fring2}, supersymmetric systems \cite{Cen}, noncommutative quantum mechanics \cite{Dey2}, etc.} Historical notes can be found e.g. in \cite{Cari, Morris}. We note that the name of the EMP equation is not yet a consensus in the literature. For instance, sometimes it can be referred just as Pinney equation. 

Diverse generalizations of the EMP equation have been proposed, as for instance allowing $C(t)$ to have a dependence on time \cite{Herring, Morris}, the inclusion of dissipation \cite{Haas2, Guha}, unbalanced systems of EMP equations with different frequency functions \cite{Ath}, modified nonlinearities \cite{Reid}, stochastic differential equations with additive noise \cite{Cer}. Following the generalization trend, it would be relevant to extend the EMP equation into the special relativity domain, with potential applications for the dynamics of charged particles in high energy density fields \cite{Hart}. To date only few results weakly related to special relativistic EMP equations are available, in connection with the Dirac equation \cite{Thi} and the relativistic isotropic two-dimensional time-dependent harmonic oscillator \cite{Petrov}. Naturally, the relativistic extension is not straightforward due to the intrinsic extra nonlinearity imposed by the presence of the Lorentz factor. Moreover, the intimate connection between the EMP equation and the time-dependent harmonic oscillator raises the question of what defines the relativistic harmonic oscillator. In this regard, the relativistic motion immediately induces anharmonicities even in the case of a quadratic potential with a frequency-amplitude dependence \cite{Guerrero, Belendez}. Here we follow the approach of many authors \cite{Edery, Poszwa, Znojil, MK, Gold, Harvey, They, Char2, Llibre2, Li}, adopting the spinless Salpeter equation \cite{Sal1} with a quadratic external potential as our definition of relativistic harmonic oscillator. The choice is due to its simplicity, where just the Newtonian kinetic energy is replaced by its special relativistic counterpart. This version of the relativistic harmonic oscillator model has been recently experimentally probed \cite{Fuji}.

The present work proposes a systematic approach towards the relativistic EMP equation and beyond, using as a simple guiding principle the Eliezer and Gray \cite{Eliezer} physical interpretation of the Ermakov-Lewis associated first integral for the isotropic two-dimensional time-dependent harmonic oscillator. The reason is that the Eliezer and Gray method based on the conservation of the angular momentum of an auxiliary planar motion provides the standard physical reasoning behind the conservation of the Ermakov invariant \cite{RayReid, Mancas}. Moreover, the extension of the Ray-Reid Ermakov systems \cite{RayReid} to the relativistic domain will be also obtained, in a certain sense to be explained. In brief, a relativistic extension of the celebrated EMP equation and extensions is proposed (which obviously has no direct relation to special solutions of Klein-Gordon, Dirac or similar relativistic partial differential equations). 

This work is organized as follows. In Section II, we review the Eliezer and Gray interpretation. In Section III, we consider the relativistic isotropic two-dimensional time-dependent harmonic oscillator and follow the Eliezer and Gray method, in order to identify a relativistic EMP equation together with the associated conservation law. From the structure of the dynamical equations, the appropriate relativistic Ray-Reid system will be also derived, as well as with the corresponding first integral. Section IV provides an alternative derivation based on a dynamical rescaling of time parameter. Sections V and VI deal with basic properties of the relativistic EMP equation, mainly in the autonomous case. Section VII is dedicated to a nonlinear superposition law relating the solutions of the relativistic EMP system. Finally, Section VIII is reserved to the conclusions and extra remarks. 

\section{The Eliezer and Gray physical interpretation}

We briefly reproduce the Eliezer and Gray physical interpretation \cite{Eliezer} of the Er\-ma\-kov-Lewis invariant, which will serve us as a guidance for a relativistic generalization of the EMP system. Consider the Lagrangian 
\begin{equation}
\label{fu}
L = \frac{1}{2}(\dot{x}^2+\dot{y}^2) - V(x,y,t) \,, \quad V(x,y,t) =\frac{\kappa^{2}(t)(x^2+y^2)}{2} 
\end{equation}
the corresponding auxiliary plane motion
\begin{equation}
\label{apm}
\ddot{\bf r} + \kappa^{2}(t){\bf r} = 0 \,,
\end{equation}
where ${\bf r}$ has the Cartesian components $(x,y)$, and the EMP equation
\begin{equation}
\label{emp1}
\ddot\rho + \kappa^2(t)\rho = \frac{J^2}{\rho^3} \,,
\end{equation}
where $J$ is a real constant. In terms of polar coordinates $(\rho,\theta)$, where $\rho = |{\bf r}|, x = \rho\cos\theta, y = \rho\sin\theta$, the equations of motion become
\begin{eqnarray}
\label{r}
\ddot\rho - \rho\dot\theta^2 + \kappa^2\rho &=& 0 \,, \\
\label{a}
\frac{1}{\rho}\frac{d}{dt}(\rho^2\dot\theta) &=& 0 \,.
\end{eqnarray}
Equation (\ref{a}) implies the constancy of the angular momentum
\begin{equation}
\label{j}
J = \rho^2\dot\theta = {\rm const.} 
\end{equation}
assuming for simplicity an unit mass. Combining Eqs. (\ref{r}) and (\ref{j}) we derive Eq. (\ref{emp1}). 
Considering the equation for the $x-$component of the auxiliary motion together with Eq. (\ref{emp1}), the constancy of the Ermakov-Lewis invariant $I$ given by 
\begin{equation}
I = \frac{1}{2}\left[(\rho\dot x - x\dot\rho)^2 + \frac{J^2 x^2}{\rho^2}\right]
\end{equation}
is directly verified, $dI/dt = 0$ along trajectories. Expressing in terms of polar coordinates, one has
\begin{equation}
I = \frac{1}{2}(J^2 \sin^{2}\theta + J^2\cos^{2}\theta) = \frac{J^2}{2} \,.
\end{equation}
Therefore, the invariance of $I$ is equivalent to the invariance of the angular momentum of the auxiliary plane motion. 

For reference, it is worth to consider the Ray-Reid (RR) generalization \cite{RayReid} of the EMP system and invariant, namely 
\begin{eqnarray}
\label{r1}
\ddot{x} + \kappa^2 x &=& \frac{f(y/x)}{y x^2} \,,\\
\label{r2}
\ddot{y} + \kappa^2 y &=& \frac{g(x/y)}{x y^2} \,,
\end{eqnarray}
where $f, g$ are arbitrary functions of the indicated arguments. 
The RR first integral for Eqs. (\ref{r1}) and (\ref{r2}) is 
\begin{equation}
\label{rri3}
I_{RR} = \frac{1}{2}(x\dot{y} - y\dot{x})^2 + \int^{y/x}f(s)ds + \int^{x/y}g(s)ds \,.
\end{equation}
One can directly verify that $dI_{RR}/dt = 0$ along trajectories. 

\section{A relativistic Ermakov-Milne-Pinney system}

Proceeding in strict analogy with the NR case, consider the equations of motion for the 2D relativistic unit rest mass time-dependent harmonic oscillator, which can be derived {\mod \cite{Goldstein, Hand}} from the Lagrangian 
\begin{equation}
\label{gl}
L = - \frac{c^2}{\gamma} - V(x,y,t) \,, \quad V(x,y,t) =\frac{\kappa^{2}(t)(x^2+y^2)}{2} \,,
\end{equation}
where $c$ is the speed of light and $\gamma = [1 - (\dot{x}^2 + \dot{y}^2)/c^2]^{-1/2}$. The Euler-Lagrange equations are 
\begin{eqnarray}
\left(1 - \frac{\dot{y}^2}{c^2}\right)\ddot{x} + \frac{\dot{x}\dot{y}}{c^2}\ddot{y} &=& - \frac{\kappa^{2}x}{\gamma^3} \,, \\ 
\left(1 - \frac{\dot{x}^2}{c^2}\right)\ddot{y} + \frac{\dot{x}\dot{y}}{c^2}\ddot{x} &=& - \frac{\kappa^{2}y}{\gamma^3} \,,
\end{eqnarray}
which can be disentangled as 
\begin{eqnarray}
\ddot x + \frac{\kappa^2}{\gamma}\frac{x}{\gamma_x^2} &=& \frac{\kappa^2}{\gamma}\frac{\dot x\dot y}{c^2}y \,, \label{x1} \\
\ddot y + \frac{\kappa^2}{\gamma}\frac{y}{\gamma_y^2} &=& \frac{\kappa^2}{\gamma}\frac{\dot x\dot y}{c^2}x \,, \label{x2}
\end{eqnarray}
where $\gamma_x = (1 - \dot{x}^2/c^2)^{-1/2},\, \gamma_y = (1 - \dot{y}^2/c^2)^{-1/2}$.
The equations of motion are in agreement with \cite{Petrov}. Similarly to the NR case, employing cylindrical coordinates $x = \rho\cos\theta, y = \rho\sin\theta$, we have the conserved angular momentum expressed as 
\begin{equation}
J = \frac{\partial L}{\partial\dot\theta} = \gamma\rho^2\dot\theta \,.
\end{equation}
Moreover, we get the Lorentz factor expressible in terms of $\rho, \dot\rho$ as 
\begin{equation}
\label{l}
\gamma = \left(\frac{1 + J^2/c^2\rho^2}{1 - \dot\rho^2/c^2}\right)^{1/2} \,.
\end{equation}
The NR formal limit $c \rightarrow \infty$ yields $\gamma = 1$ also from Eq. (\ref{l}). 
Using the angular momentum to eliminate the angular velocity, we obtain 
\begin{eqnarray}
\label{e1}
\ddot{x} + \frac{\kappa^2}{\gamma}\left(x - \frac{\rho\dot\rho\dot x}{c^2}\right) &=& 0 \,,\\
\label{e2}
\ddot\rho + \frac{\kappa^2}{\gamma}\left(1 - \frac{\dot\rho^2}{c^2}\right)\rho &=& \frac{J^2}{\gamma^2\rho^3} \,.
\end{eqnarray}
From the identity $\gamma^2(\rho\dot x - x\dot\rho)^2 = J^2 \sin^{2}\theta$, it becomes self-evident that the quantity 
\begin{equation}
\label{e3}
I_R = \frac{1}{2}\left[\gamma^2(\rho\dot x - x\dot\rho)^2 + \frac{J^2 x^2}{\rho^2}\right]
\end{equation}
is a first integral, since 
\begin{equation}
I_R = \frac{1}{2}(J^2\sin^{2}\theta + J^2\cos^{2}\theta) = \frac{J^2}{2} \,,
\end{equation}
in complete analogy with the Ermakov-Lewis invariant for the NR problem. It can be also directly verified that $dI_R/dt = 0$ along the trajectories of the system (\ref{e1})-(\ref{e2}). The invariant has the same physical interpretation of the NR Ermakov invariant in terms of the conservation of the angular motion of the associated auxiliary 2D motion. In this context, it is justified to interpret Eqs. (\ref{e1}) and (\ref{e2}) as the (special) relativistic EMP system. Equation (\ref{e2}) is a relativistic EMP equation (REMP), and the first integral in Eq. (\ref{e3}) is the relativistic Ermakov-Lewis invariant of the problem. Obviously, in the formal limit $c \rightarrow \infty$ one recovers the NR case. Understanding the Lorentz factor in the sense of Eq. (\ref{l}), the relativistic EMP system is a pair of nonlinear second-order ordinary differential equations for $x, \rho$. Unlike the NR case, the equation for $x$ is not uncoupled. 

{ To recapitulate, we have just used the standard Lagrangian for the special relativistic two-dimensional motion \cite{Landau, Jackson, Corben, Arya}, in the particular case where the particle is under the influence of a force linear in position (a quadratic potential) as seen in a fixed local inertial frame with respect to which the motion is observed \cite{Goldstein, Hand}.}

{ In passing, we write the Hamiltonian ${\cal H}$ associated with (\ref{gl}), 
\begin{equation}
{\cal H} = c^2 \left(1 + \frac{p_x^2 + p_y^2}{c^2}\right)^{1/2} + V(x,y,t) \,, \quad p_{x} = \gamma\dot x \,,\,\, p_y = \gamma\dot y \,,
\end{equation}
which is not a constant of motion in the non-conservative case where $\kappa(t)$ is a time-dependent function. Even in the autonomous case where $d{\cal H}/dt = 0$, the nature of ${\cal H}$ and the invariant $I_R$ in Eq. (\ref{e3}) is different: the former is energy-like, while the latter is angular momentum-like.}

From the structure of Eqs. (\ref{e1}) and (\ref{e2}) and after some trial and error, it is possible to identify a relativistic Ray-Reid (RRR) system, namely
\begin{eqnarray}
\label{rr1}
\ddot{x} + \frac{\kappa^2}{\gamma}\left(x - \frac{\rho\dot\rho\dot x}{c^2}\right) &=& \left(1 - \frac{\dot{x}^2}{c^2}\right)\frac{f(y/x)}{\gamma^2 y x^2} - \frac{\dot{x}\dot{y}}{\gamma^2 c^2}\frac{g(x/y)}{x y^2}    \,,\\
\label{rr2}
\ddot{y} + \frac{\kappa^2}{\gamma}\left(y - \frac{\rho\dot\rho\dot y}{c^2}\right) &=& - \,\frac{\dot{x}\dot{y}}{\gamma^2 c^2}\frac{f(y/x)}{y x^2}  + \left(1 - \frac{\dot{y}^2}{c^2}\right)   \frac{g(x/y)}{\gamma^2 x y^2} \,,
\end{eqnarray}
where $\rho\dot\rho = x\dot x + y\dot y$ and the Lorentz factor is understood as a function of $\dot{x}, \dot{y}$. The invariant for Eqs. (\ref{rr1}) and (\ref{rr2}) is 
\begin{equation}
\label{rr3}
I_{RRR} = \frac{\gamma^2}{2}(x\dot{y} - y\dot{x})^2 + \int^{y/x}f(s)ds + \int^{x/y}g(s)ds \,.
\end{equation}
It can be verified that $dI_{RRR}/dt = 0$ along trajectories. It is apparent that Eqs. (\ref{rr1}), (\ref{rr2}) and (\ref{rr3}) define  a RRR system and its invariant, showing a complete symmetry between the $x$ and $y$ variables and recovering the RR system and invariant in the formal NR limit $c \rightarrow \infty$, as shown by comparison with Eqs. (\ref{r1})-(\ref{rri3}).  
A derivation provided by a dynamical rescaling of time will be described in the next Section.

Although our treatment has as motivation the relativistic time-dependent harmonic oscillator, it is clear that the invariants shown in Eqs. (\ref{e3}) and (\ref{rr3}) do not depend on $\kappa$. Therefore, one is authorized to allow for more general functional dependencies of $\kappa$, e.g. $\kappa = \kappa(t,x,y,\dot{x},\dot{y},\dots)$ in Eqs. (\ref{rr1}) and (\ref{rr2}), maintaining the constancy of the RRR invariant. Similar remarks apply to the NR case \cite{ReidRay, HG, GH}. { Notice that Eqs. (\ref{rr1}) and (\ref{rr2}) do not have a Lagrangian structure in general.}

\section{Derivation from a dynamical rescaling of time}

Start from the dynamical system 
\begin{equation}
\label{rrr1}
x'' + \omega^2 x = \frac{f(y/x)}{y x^2} \,, \quad y'' + \omega^2 y = \frac{g(x/y)}{x y^2} \,,
\end{equation}
where a prime denotes derivative with respect to the independent variable $\tau$, or $x' = dx/d\tau, y' = dy/d\tau$, and where $\omega$ can be anything in the same spirit of the last Section. In the same way as for the RR system, it is obvious that (\ref{rrr1}) possess the invariant 
\begin{equation}
I = \frac{1}{2}(x y' - y x')^2 + \int^{y/x}f(s)ds + \int^{x/y}g(s)ds \,, \quad \frac{dI}{d\tau} = 0 \,. \label{kkkk}
\end{equation}
If we now move to a new independent variable $t$ defined by 
\begin{equation}
\label{drt}
\frac{dt}{d\tau} = \gamma \,, \quad \gamma = \left(1 - \frac{\dot{x}^2+\dot{y}^2}{c^2}\right)^{-1/2} \,, \quad \dot{x} = \frac{dx}{dt} \,, \quad \dot{y} = \frac{dy}{dt} \,,
\end{equation}
then it can be checked after some algebra that the RRR system (\ref{rr1}) and (\ref{rr2}) is recovered, with $\kappa^2 = \omega^2/\gamma$, and that the invariant (\ref{kkkk}) transforms into (\ref{rr3}). The procedure gives a more transparent derivation of the RRR system from a dynamical rescaling of time starting from the RR system. 

{ It is worthwhile to have a discussion about Lorentz invariance. In the same sense as the Newtonian harmonic oscillator is obviously not Galilean invariant, the relativistic harmonic oscillator is not Lorentz invariant. In this respect we remember that the symmetry group of the NR Ermakov system is the $SL(2,\Re)$ group \cite{Leach}, which has no relation with the Galilean group. 
Moreover notice that the force equation $dp_i/dt = F_i$ where $p_i$ are the components of the relativistic momentum is relativistic only ``in a certain sense", quoting the words from Golsdtein's book \cite{Goldstein}, since time has been keep as entirely distinct from the spatial coordinates. 
{ However, a fully covariant interpretation is possible, regarding (\ref{gl}) as the Lagrangian for a charged particle under a scalar electrostatic potential which is a quadratic function of the spatial coordinates, in the instant inertial frame with respect the motion is observed. In this context, a fully covariant formalism could be easily performed following the standard approach \cite{Goldstein, Hand, Landau, Jackson, Corben, Arya} starting with the Lagrangian $L = - mc^2/\gamma + q {\bf A}\cdot{\bf r} - q\phi$ (the well-known notation is employed) in the two-spatial dimensions case, with a scalar potential $\phi \sim (x^2 + y^2)$ together with ${\bf A} = 0$. This is the Lagrangian (\ref{gl}) assuming unit rest mass without loss of generality. 
In addition, }the relativistic harmonic oscillator and REMP equations reproduce the Newtonian systems in the limit { $u/c \rightarrow 0$}, where $u$ is a measure of the maximal velocity of the problem. For instance if $A_0$ is the amplitude of the motion under a linear force $F = - \kappa^2 x$, the relativistic effects become negligible provided $\kappa A_0/c \rightarrow 0$. The case of a very strong external field acting on a charged particle [29] is a suitable system to probe the relativistic effects. In this context since the non-relativistic Ermakov system has found applications in accelerator physics \cite{Qin}, the relativistic 
version has potential applications for charged particle motions under high-intensity external fields.

Notice that the derivation of the equations (\ref{e1}) and (\ref{e2}) containing the REMP equation, together with the relativistic Ermakov-Lewis invariant (\ref{e3}), follows a very different route compared to the RRR system (\ref{rr1})-(\ref{rr2}) and associated invariant (\ref{rr3}). To obtain the REMP equation, the starting point was the relativistic 
Lagrangian 
for the 2D time-dependent harmonic oscillator, together with polar coordinates, exactly the same as the procedure for the NR EMP system but then with the special relativistic Lagrangian. On the other hand, the RRR system can be considered ``relativistic" only in the {\it ad hoc} sense that it reduces to the NR Ray-Reid system in the limit $c \rightarrow \infty$ where $c$ is some reference velocity, since its derivation does not starts from any physical relativistic Lagrangian. In the same context, the use of a dynamical rescaling of time clearly has not a relativistic spirit. First of all, the $\gamma$ factor in Eq. (\ref{drt}) is defined in terms of the particle velocity, not the relative velocity between two inertial frames. Second, it is obviously not a Lorentz transformation since the space variables are kept the same. Third, we had not the objective of setting a sort of symmetry transformation for the equations (\ref{rrr1}). Rather, the goal was the use of a non-invariance transformation towards the derivation of the RRR system. Although at the moment there is the lacking of a a physical interpretation of the RRR system, which as discussed is relativistic only in a formal sense, we think it is worthwhile to present it since it has a striking analogy with the traditional RR system, being a non-trivial pair of coupled nonlinear second-order ordinary differential equations possessing an exact invariant. }

\section{The $J = 0$ case}

For a vanishing angular momentum, one has $\dot\theta = 0$ so that it can be chosen $\theta = 0$, without loss of generality. In this case, $y = 0$ and Eq. (\ref{e1}) becomes
\begin{equation}
\label{oned}
\frac{d}{dt}(\gamma \dot x) = - \kappa^2 x \,, \quad \gamma = \left(1-\frac{\dot x^2}{c^2}\right)^{-1/2} \,,
\end{equation}
which is \cite{Bouquet} the equation for an one-dimensional (1D) relativistic time-dependent harmonic oscillator. 

\begin{figure}[h]
\begin{center}
\includegraphics[width=4.5in]{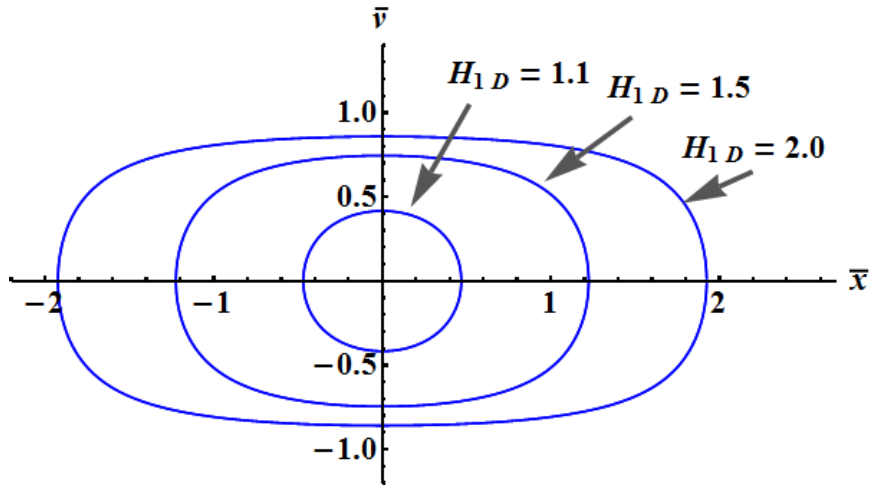}
\caption{Phase-space contour plots of the energy first integral (\ref{ham}) for the  1D conservative relativistic harmonic oscillator described by Eq. (\ref{oned}) with constant $\kappa$, for $\bar x = \kappa x/c,\,\bar{v} =\dot x/c$ and different values of $H_{1D}$, as indicated.}
\label{fig1} 
\end{center}
\end{figure}

In the case of a constant frequency $\kappa$, obviously the energy is conserved. In terms of rescaled variables $\bar{x} = \kappa x/c,\, \bar{t} = \kappa t,\, \bar{v}=d\bar{x}/d\bar{t}$, the corresponding first integral $H_{1D} \geq 1$ is
\begin{equation}
\label{ham}
H_{1D} = (1-\bar{v}^2)^{-1/2}+\frac{\bar{x}^2}{2} \,,
\end{equation}
with phase-space contour plots shown in Fig. \ref{fig1}. Clearly, only bounded trajectories are admissible, as expected. The return points are located at $\bar{x} = \pm \sqrt{2}(H_{1D}-1)^{1/2}$. In rescaled coordinates, the NR limit $H_{1D} \simeq 1$ corresponds to circular trajectories in phase space, { $\bar{v}^2 + \bar{x}^2 \simeq 2(H_{1D}-1)$.} For larger values of $H_{1D}$, the return points are also larger, implying more relativistic effects and increasing anharmonicity, as seen in Fig. \ref{fig1}. 

\begin{figure}[ht]
\begin{center}
\includegraphics[width=4.5in]{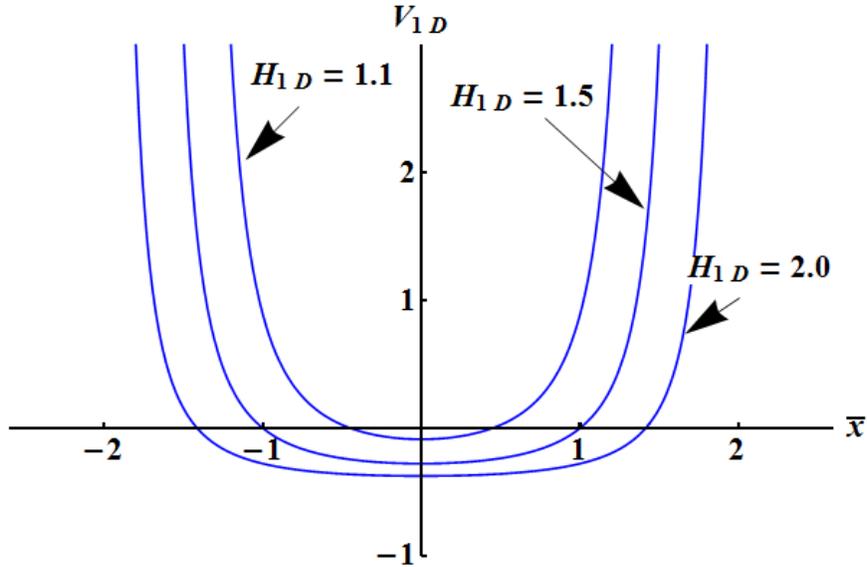}
\caption{Pseudo-potential $V_{1D}(\bar{x})$ from Eq. (\ref{vv1d}) and different values of $H_{1D}$, as indicated.}
\label{fig2} 
\end{center}
\end{figure}

It is instructive to rewrite the conservation law in a Newtonian form, 
\begin{equation}
\label{v1d}
\frac{\bar{v}^2}{2} + V_{1D}(\bar{x}) = 0 \,,
\end{equation}
where the pseudo-potential $V_{1D}(\bar{x})$ is defined by
\begin{equation}
\label{vv1d}
V_{1D}(\bar{x}) = -\,\frac{1}{2} + \frac{1}{2(H_{1D} - \bar{x}^2/2)^2} \,,
\end{equation}
shown in Fig. \ref{fig2}. In the NR limit, the variable $\bar{x}$ is limited to small values, so that $V_{1D} = {\rm cte.} + \bar{x}^2/(2H_{1D}^3) + \dots$ with $H_{1D} \simeq 1$, while larger values of $H_{1D}$ correspond to enhanced relativistic effects and anharmonicity. { The quantity $V_{1D}$ provides a kind of transfer of the nonlinearity from the kinetic energy term to an anharmonic (pseudo) potential term, but seemingly has not a physical interpretation.} 

The conserved energy can be used for the quadrature of the motion in terms of elliptic functions. 
{ For this purpose, we set 
\begin{equation}
H_{1D} = 1 + \frac{A^2}{2} \,, \quad - A < \bar{x} < A \,,
\end{equation}
where $A$ is the amplitude of the motion, and define 
\begin{equation}
\bar{x} = A\sin\phi \,, \quad \phi = \phi(\bar{t}) \,.
\end{equation}
After some algebra, Eq. (\ref{v1d}) is rewritten as 
\begin{equation}
\left(\frac{d\phi}{d\bar{t}}\right)^2 = \frac{1 + \frac{A^2}{4}\cos^2\phi}{(1 + \frac{A^2}{2}\cos^2\phi)^2} \,,
\end{equation}
which can be immediately integrated yielding 
\begin{equation}
\sqrt{4+A^2}\,E\left(\phi, \frac{A}{\sqrt{4+A^2}}\right) - \frac{2}{\sqrt{4+A^2}}\,F\left(\phi, \frac{A}{\sqrt{4+A^2}}\right) = \bar{t}-\bar{t_0} \,,
\end{equation}
where $\bar{t}_0$ is an integration constant and $F, E$ are incomplete elliptic integrals of the first and second kind, respectively defined \cite{MilneT} according to 
\begin{equation}
F(\phi,k) = \int_{0}^{\phi} \frac{d\phi'}{(1 - k^2 \sin^2\phi')^{1/2}} \,, \quad E(\phi,k) = \int_{0}^{\phi} (1 - k^2 \sin^2\phi')^{1/2}d\phi' \,.
\end{equation}

Using MATHEMATICA it is easy to produce a series solution, 
\begin{equation}
\phi(\bar{t}) = \bar{t}-\bar{t}_0 - \frac{3A^2}{32}(2(\bar{t}-\bar{t}_0) + \sin[2(\bar{t}-\bar{t}_0)]) + {\cal O}(A^4) \,,
\end{equation}
recovering the Newtonian result in the limit of small amplitude (which is also the non-relativistic limit). Similarly the period $T$ for which $\phi=2\pi$ is 
\begin{equation}
T = 2\pi\left(1 + \frac{3A^2}{16}\right) + {\cal O}(A^4) \,.
\end{equation}
Approximate periodic solutions for the 1D conservative relativistic harmonic oscillator can also be found in \cite{Mickens}.}

\section{Relativistic conservative Ermakov-Milne-Pinney equation}

The REMP defined in Eq. (\ref{e2}) with $J \neq 0$ does not have collapsing ($\rho \rightarrow 0$) solutions due to the inverse cubic term. For instance, suppose $\kappa = {\rm cte.}$ and the rescaling $\bar\rho = \kappa\rho/c, \bar{t} = \kappa t, \bar{v} = d\bar\rho/d\bar{t}, \bar{J} = \kappa J/c^2$, so that
\begin{equation}
\frac{d\bar{v}}{d\bar{t}} + \frac{1}{\gamma}(1-\bar{v}^2)\bar\rho = \frac{\bar{J}^2}{\gamma^2\bar{\rho}^3} \,.
\end{equation}
The energy first integral is
\begin{equation}
\label{h} 
H = \gamma + \frac{\bar\rho^2}{2}  \geq 1  \,, \quad \gamma = \left(\frac{1 + \bar{J}^2/\bar\rho^2}{1 - \bar{v}^2}\right)^{1/2} \,.
\end{equation}
It is immediate to conclude that $0 < \bar\rho < \sqrt{2H}$. 

\begin{figure}[ht]
\begin{center}
\includegraphics[width=8.0cm,height=6.0cm]{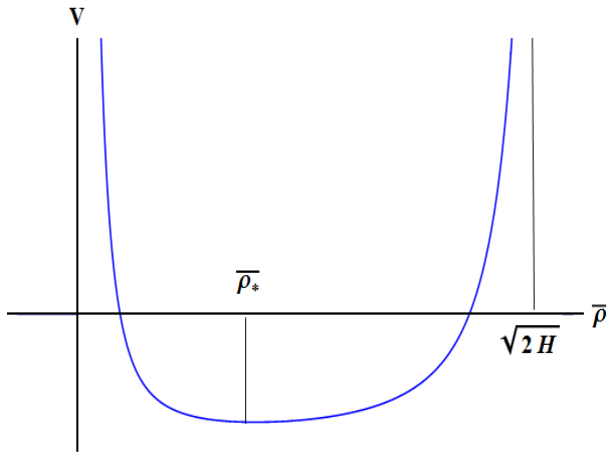}
\end{center}
\caption{Pseudo-potential $V$ from Eq. (\ref{vv}).} 
\label{fig3}
\end{figure}

Rewriting in a Newtonian form yields 
\begin{equation}
\label{red}
\bar{v}^2/2 + V(\bar\rho) = 0 \,,
\end{equation}
with a pseudo-potential defined by
\begin{equation}
\label{vv}
V(\bar\rho) = - \frac{1}{2} + \frac{1+\bar{J}^2/\bar\rho^2}{2(H - \bar\rho^2/2)^2} \,,
\end{equation}
shown in Fig. \ref{fig3}. It is possible to show that $V(\bar\rho_*) < 0$, where $0 < \bar\rho_* < 2H$ is the equilibrium point of $V$, viz.
\begin{equation}
\bar\rho_* = \frac{1}{2}\left(-3\bar{J}^2 + \sqrt{9\bar{J}^4 + 16\bar{J}^2 H}\right)^{1/2} \,.
\end{equation}

\begin{figure}[ht]
\begin{center}
\includegraphics[width=8.0cm,height=6.0cm]{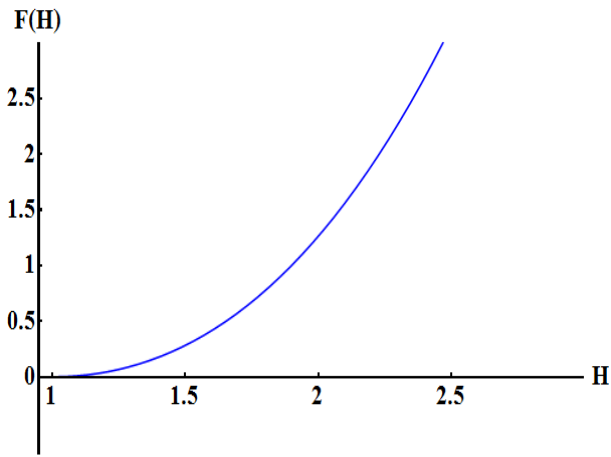}
\end{center}
\caption{Characteristic function $F(H)$ from Eq. (\ref{ff}) for $H \geq 1$.}
\label{fig4}
\end{figure}

The return points correspond to $V = 0$, similarly to Fig. \ref{fig2}, with stable oscillations around $\bar\rho_*$.
Elementary algebra shows that the condition for periodic motions ($V(\bar\rho_*) < 0$) is equivalent to 
\begin{equation}
\label{ff}
\bar{J}^2 < F(H) = \frac{4}{27}\left(-9H+H^3+ (3+H^2)^{3/2}\right) \,,
\end{equation}
or, sufficiently small angular momentum. In the NR limit one has ${\bar J}^2 < (H-1)^2$ disregarding ${\cal O}(H-1)^3$ terms, so that the periodicity condition can be shown to be automatically satisfied. The characteristic function $F$ is shown in Fig. \ref{fig4}. A numerical investigation shows that the periodicity condition is always meet, for meaningful initial conditions $0 < \bar\rho(0) < \sqrt{2H},\, -1 < \bar{v}(0) < 1$, as expected. 

The quadrature of Eq. (\ref{red}) can be made in terms of elliptic functions, after determining the return points corresponding to the potential well from Eq. (\ref{vv}). However, the result is exceedingly complicated.

\section{Nonlinear superposition law}

Suppose $\rho = \rho(t)$ a particular solution of the REMP equation (\ref{e2}). Introducing the new variables 
\begin{equation}
Q = \frac{x}{\rho} \,, \quad T = \int\frac{dt}{\gamma\rho^2} \,,
\end{equation}
where $\gamma = \gamma(t)$ is given by Eq. (\ref{l}) converts the relativistic Ermakov-Lewis invariant (\ref{e3}) into
\begin{equation}
I_R = \frac{1}{2}\left(\frac{dQ}{dT}\right)^2 + \frac{J^2 Q^2}{2} \,,
\end{equation}
formally the same as the energy first integral for a 1D conservative NR harmonic oscillator. A quadrature yields
\begin{equation}
Q = \frac{\sqrt{2 I_R}}{|J|}\sin(J T + \delta) \,,
\end{equation}
or
\begin{equation}
\label{nsl}
x = \rho\sin\left(J\int\frac{dt}{\gamma\rho^2} + \delta\right) \,,
\end{equation}
since $I_R = J^2/2$, adopting $J > 0$ and where $\delta$ is a constant phase. The nonlinear superposition law (\ref{nsl}) generalizes the Newtonian result \cite{ReidRay} to the relativistic context. In concrete applications, typically the particular solution $\rho$ should be  numerically found.

\section{Conclusion}

In this work, considerable progress was achieved, in the generalization of Ermakov systems towards the special relativity domain. The Eliezer and Gray physical interpretation of the Ermakov-Lewis invariant, was used as a guide for the derivation of the relativistic analog of the EMP equation, together with the corresponding first integral for the relativistic planar time-dependent harmonic oscillator. General aspects of the relativistic EMP equation have been addressed, and a nonlinear superposition law was derived. 
In spite of the successful results, it is still possible to derive other classes of relativistic Ermakov systems, not arising from the correspondence with the Eliezer and Gray physical interpretation. For instance, symmetry principles can be a guiding principle, although the $SL(2,\Re)$ group structure of non-relativistic Ermakov systems obviously tends to be broken in the relativistic domain. In the same footing, to carry to the relativistic domain the linearization of standard Ermakov systems would be a probably unfeasible task. 
{ In future works it is worthwhile to see how
one can actually solve the new relativistic Ermakov systems with an explicit time-dependence. For this purpose the search for quasi-exact solutions \cite{fring5} can be a fruitful approach.} 

\acknowledgments
This research was funded by Con\-se\-lho Na\-cio\-nal de De\-sen\-vol\-vi\-men\-to Cien\-t\'{\i}\-fi\-co e Tec\-no\-l\'o\-gi\-co (CNPq).



\end{document}